\documentclass[12pt]{article}
\input epsf.tex
\epsfclipon
\usepackage{epsfig}
\usepackage{epstopdf}
\usepackage{epsf}
\usepackage{graphicx,amsmath,amssymb}
\usepackage{epsfig,multicol}
 \usepackage{epsfig}
\usepackage{amssymb}
\usepackage{jheppub}
\def\bg{\begin{eqnarray}}
\def\nd{\end{eqnarray}}

\begin{document}
\title{Universal Bounds on Operator Dimensions in General 2D Conformal Field Theories }
\author{Joshua D. Qualls}
\affiliation{Department of Physics, National Taiwan University, Taipei, Taiwan}
\emailAdd{joshqualls@ntu.edu.tw}

\abstract{
We derive a bound on the conformal dimensions of the lightest few states in general unitary 2d conformal field theories with discrete spectra using modular invariance, including CFTs with chiral currents. We derive a bound on the conformal dimensions $\Delta_1$ and $\Delta_2$  going as $c_{\rm tot}/12 + O(1)$. The bound is of the same form found for CFTs without chiral currents in \cite{hell} and \cite{paper1}. We then prove the inequality $\Delta_n \leq c_{\rm tot}/12 + O(1)$ for large $c_{\rm tot}$ and with appropriate restrictions on $n$. Using the AdS$_3$/CFT$_2$ correspondence, our bounds correspond to upper bounds on the masses of the lightest few states and a lower bound on the number of states. We conclude by checking our results against several candidate conformal field theories.
}
\maketitle

\section{Introduction}

Recent years have seen a resurgence in constructing conformal field theories (CFTs) from consistency conditions imposed by conformal invariance--the so-called conformal bootstrap program \cite{u1,u2,u3}. In two dimensions with central charge $c<1$, local conformal symmetry has given particularly powerful constraints \cite{u4,blumen}. Relatively little is known, however, about the general landscape of CFTs in higher spacetime dimensions. More recently, a number of broad constraints on the spectra and structure constants of CFTs have been obtained by considering operator product expansion (OPE) associativity of correlators \cite{h1,h2,h3,h4,h5,h6,h7,h8,h9,h10,h11,h12,h13,h14,h15,h16,h17,h18,h181,h182,h183,h184}. One finds even more powerful constraints in the presence of supersymmetry \cite{h191,h192,h20,h21,h22,h23,h24,h25,h26,h27,h28} or by studying  large-spin asymptotics of the operator spectrum in the lightcone limit \cite{h29,h30,h31,h32}.

These OPE associativity techniques can still be applied to two-dimensional CFTs. Local conformal symmetry in two dimensions is special, however, and can supply additional powerful constraints. One example of this comes from demanding consistency of a CFT on arbitrary Riemann surfaces. In two dimensions, crossing symmetry of the four-point functions on the sphere and modular invariance of the partition function and one-point functions on the torus are necessary and sufficient conditions for the theory to be consistently defined on all two-dimensional surfaces \cite{m1}. We are therefore interested in constraints coming from modular invariance, as this approach can be understood as somewhat complementary to OPE associativity techniques.

In \cite{hell}, Hellerman used modular invariance of the partition function to derive a bound on $\Delta_1 = h+\bar{h} $, the conformal dimension of the lowest nonvacuum primary operator in terms of the right- and left-moving central charges $c$ and $\bar{c}$:
\begin{equation}
	\Delta_1 \leq \frac{c+\tilde{c}}{12} +  \frac{(12-\pi) + (13\pi -12)e^{-2\pi}}{6\pi(1-e^{-2\pi})}  \equiv   \frac{c_{\rm tot}}{12} + \delta_0, \,\,\,\, \delta_0 \approx  0.4736... \label{eq:hellbound}
\end{equation}
This bound holds for any unitary 2d CFT having left and right central charge $c, \tilde{c}>1$ and with no chiral primary operators other than the chiral components of the stress tensor (and their chiral descendants). By chiral primary operators, we mean chiral operators with respect to the Virasoro algebra---operators having $h=0,\bar{h}>0$ or vice versa. Building on this work, higher order modular invariance constraints were used in \cite{fried} to find that at finite $c_{\rm tot}$ the bound can be lowered significantly (while for large $c_{\rm tot}$ the bounds apparently asymptote to $\frac{c_{\rm tot}}{12}$). In \cite{hart}, modular invariance was used to bound the number of states in a given range of energies subject to some conditions on the spectrum. Modular invariance was used in \cite{hell2} to bound the number of operators. Modular constraints on 2d CFT spectra were connected to Calabi-Yau compactifications in \cite{kell}.

Of course, additional assumptions on the 2d CFT lead to tighter bounds on $\Delta_1$. 
For example, \cite{6p} (see also \cite{9p,10p} examined 2d CFTs for which the partition function is holomorphically factorized as a function of the complex structure $\tau$ of the torus. In this class of CFT, it can be shown that the lowest primary operator is either purely left- or right-moving, and can have a weight no larger than $1 + \mbox{min}(\frac{c}{24},\frac{\tilde{c}}{24})$. Other work \cite{11p,fried} considers  a certain subclass of (2,2) SUSY CFTs that suggest a bound that goes as $\Delta_1\leq\frac{c}{24}$ for large central charge. In \cite{paper2}, a similar bound was obtained for modular invariant 2d CFTs having only even-spin primary operators.

In this note, we move the opposite direction and extend the arguments of \cite{hell} and \cite{paper1} to derive an analytic bound on the conformal dimensions $\Delta_1$ and $\Delta_2$ for any unitary 2d CFT with discrete spectra and with left and right central charge $c, \tilde{c}>1$---that is, we remove the restriction that there are no chiral primary operators other than the chiral components of the stress tensor and their descendants. The bounds we obtain take the same form as Hellerman's bound (\ref{eq:hellbound}), with the same asymptotic growth $c_{\rm tot}/12$. We also investigate the possibility of deriving bounds on primary operator conformal dimensions $\Delta_n$ for $n>2$. We find that in order to derive bounds for $\Delta_n$, we need to assume a larger minimum value for $c_{\rm tot}$ that grows logarithmically with $n$.  For large enough $c_{\rm tot}$ with appropriate conditions on $n$, we show that 
$$
\Delta_n \le \frac{c_{\rm tot}}{12}+O(1).
$$

These results also have a gravitational interpretation; using the AdS$_3$/CFT$_2$ correspondence, our bounds correspond to upper bounds on the masses of the lightest few states:
\begin{equation}
M_{1,2} \leq \frac{1}{4G_N} + O(L^{-1}),
\end{equation}
where $G_N$ is Newton's constant and $L$ is the AdS radius.
These results hold for chiral theories; in particular, they apply to theories of 3d gravity coupled to chiral matter and gauge fields. The extension of this proof to CFTs with chiral Virasoro primaries means that we can apply our results a much larger class of three-dimensional spacetimes with negative cosmological constant.


\section{A bound on $\Delta_1$}

\subsection{Setup}
We begin by extending the methods of \cite{hell}. Consider a 2d CFT on the torus with its modular parameter close to the fixed point of the $S$-transformation 
$$
\tau\equiv\ (\mathcal{K}+i\beta)/2\pi=i,
$$ 
where $\beta$ is the inverse temperature and $\mathcal{K}$ is the thermodynamic potential for spatial momentum in the compact spatial direction $\sigma_1$. For purely imaginary $\tau$, the $S$-invariance of the partition function can be expressed as 
\begin{equation}
Z(\beta) = Z\left(\frac{4\pi^2}{\beta}\right).
\end{equation}
By taking successive derivatives of this expression, one obtains an infinite set of differential constraints on the partition function
\begin{equation}
\left(\beta\frac{\partial}{\partial \beta}\right)^{N}   Z(\beta)\bigg|_{\beta=2\pi}=0, \;\;N \text{ odd} \label{eq:twothreeish}. \end{equation}

We next consider the partition function of a CFT with right- and left-moving central charges $c$ and $\bar{c}$ in terms of Virasoro characters \cite{fried}:
\begin{equation}
Z(\tau,\bar{\tau}) = |\eta(\tau)|^{-2}\sum_{(h,\bar{h})\in S} N_{\bar{h}h}  \overline{ \hat{\chi}_{  \bar{h}  }  (\tau)  }   \hat{ \chi}_{  {h}  }(\tau).
\end{equation}
Here $N_{\bar{h}h}$ is the number of primary operators with conformal weights $(h, \bar{h})$, and the characters are given by the expressions
\begin{equation}
 \overline{ \hat{\chi}_{  \bar{h}  }  (\tau)  }   \hat{ \chi}_{  {h}  }(\tau) = \left\{ \begin{array}{rr}
\bar{q}^{-\frac{\tilde{c}-1}{24}}(1-\bar{q})q^{-\frac{{c}-1}{24}}(1-q),   &\;\;\;\;\; \bar{h}=0,h=0  \\
\bar{q}^{\bar{h}-\frac{\tilde{c}-1}{24}}q^{h-\frac{{c}-1}{24}}(1-q),     & \;\;\;\;\;\bar{h}>0,h=0 \\
\bar{q}^{\bar{h}-\frac{\tilde{c}-1}{24}}(1-\bar{q})q^{h-\frac{{c}-1}{24}},     &\;\;\;\;\; \bar{h}=0,h>0 \\
\bar{q}^{\bar{h}-\frac{\tilde{c}-1}{24}}q^{h-\frac{{c}-1}{24}},     &\;\;\;\;\; \bar{h}>0,h>0
\end{array} \right.
\end{equation}
We can simplify these expressions a great deal. First, we find it useful to define the shifted vacuum energy $\hat{E}_0 \equiv \frac{1}{12}+E_0=\frac{1}{12}-\frac{c_{\rm tot}}{24}$.  We also use the fact that for purely imaginary $\tau = i \beta / 2\pi$, the variable $q=\exp(2\pi i \tau) = \exp(-\beta)=\bar{q}$. Finally, instead of conformal weights we can use conformal dimension $\Delta=h+\bar{h}$ to find
\begin{equation}
 \overline{ \hat{\chi}_{  \bar{h}  }  (\tau)  }   \hat{ \chi}_{  {h}  }(\tau) = \left\{ \begin{array}{rl}
e^{-\beta \hat{E}_0}   (1-e^{-\beta})^2, &\;\;\;\;\;\bar{h}=0,h=0  \\
e^{-\beta (\Delta+\hat{E}_0)}  (1-e^{-\beta}),  &\;\;\;\;\;\bar{h}>0,h=0 \\
e^{-\beta (\Delta+\hat{E}_0)}   (1-e^{-\beta}), &\;\;\;\;\;\bar{h}=0,h>0 \\
e^{-\beta (\Delta+\hat{E}_0)},   &\;\;\;\;\;\bar{h}>0,h>0
\end{array} \right.
\end{equation}

Let us arrange the conformal dimensions in increasing order $0 < \Delta_1 \leq \Delta_2 \leq ..$. Then we can express the partition function in terms of Virasoro primaries as the sum of a vacuum contribution and non-vacuum contributions:
\begin{equation}
Z(\beta) = Z_A(\beta) +  Z_0(\beta), \label{eq:twofiveish}
\end{equation}
$$
Z_A(\beta) \equiv  |\eta( i \beta / 2\pi)|^{-2}\sum_{A=1}^{\infty} e^{-\beta(\Delta_A+\hat{E}_0)} (1-e^{-\beta})^{\delta_{h_A 0} +\delta_{\bar{h}_A 0}},
$$
$$
Z_0(\beta) \equiv |\eta( i \beta / 2\pi)|^{-2} e^{-\beta \hat{E}_0}(1-e^{-\beta})^2.
$$
We have separated the unique vacuum contribution with $h=\bar{h}=0$; the first term is the sum over conformal weights with one or both conformal weights being nonzero.

We can now apply the differential constraints (\ref{eq:twothreeish}) to the partition function (\ref{eq:twofiveish}). Following \cite{hell}, we introduce polynomials $f_p(z)$ defined by
\begin{equation}
(\beta \partial_\beta)^{p}Z_A(\beta)\bigg|_{\beta=2\pi} = \frac{(-1)^{p} e^{-2\pi \hat{E}_0}}{\eta(i)^{2}}
\sum_{A=1}^\infty e^{-2\pi \Delta_A}f_p(\Delta_A+\hat{E}_0)(1-e^{-2\pi})^{\delta_{h_A 0} +\delta_{\bar{h}_A 0}}.
\end{equation}
Although we have expressed the polynomials $f_p$ as functions of $\Delta_A$, they are in fact functions of $h_A$ and $\bar{h}_A$. In the case with no other chiral operators, this distinction was unnecessary. In the current case, however, we are interested in deriving bounds on $\Delta_A$, and explicit dependence on $h_A$ or $\bar{h}_A$ additionally shows up in Kronecker deltas multiplying new terms. We simply note and remember that there will be some additional dependence on $h_A,\bar{h}_A$ that is on occasion suppressed. The first few poynomials are explicitly
$$f_0(z)=1 $$
\begin{equation} f_1(z)=(2\pi z)-\frac12 -\frac{2\pi}{  (e^{2\pi}-1)} (\delta_{h_A 0} +\delta_{\bar{h}_A 0})    \label{eq:fs} \end{equation}
$$f_2(z)=(2\pi z)^2  -  (2\pi z)  \left(2 +  \frac{4\pi}{ e^{2\pi}-1} (\delta_{0h} +\delta_{0\bar{h}})  \right)+\left(\frac{7}{8}+2r_{20}\right) $$
$$- 4\pi\left( \frac{\pi e^{2\pi} - e^{2\pi}  + 1   } {(e^{2\pi}  - 1)^2}   \right)(\delta_{0h} +\delta_{0\bar{h}})    +  \frac{4\pi^2  } {(e^{2\pi}-1)^2} (\delta_{0h} +\delta_{0\bar{h}})^2      $$

We likewise define the polynomials $b_p(z)$ by
\begin{equation}
(\beta \partial_\beta)^{p}Z_0(\beta)\bigg|_{\beta=2\pi} = (-1)^{p} \eta(i)^{-2}\text{exp}(-2\pi \hat{E}_0)b_p(\hat{E}_0),
\end{equation}
Explicitly, we can express these polynomials as
\begin{equation}
b_p(z)= f_p(z) - 2e^{-2\pi} f_p(z+1) + e^{-4\pi}f_p(z+2)\bigg| _{h, \bar{h}>0} . \label{eq:bs}
\end{equation}
Using these polynomials, the equations (\ref{eq:twothreeish}) for modular invariance of $Z(\beta)$ for odd $p$ can be expressed as
\begin{equation}
\sum_{A=1}^{\infty}f_p(\Delta_A+\hat{E}_0)    (1-e^{-2\pi})^{\delta_{h_A 0} +\delta_{\bar{h}_A 0}}
\text{exp}(-2\pi\Delta_A)=-b_p(\hat{E}_0)  \label{eq:modeq}
\end{equation}
To simplify some expressions, we will also define the quantity
$$
(1-e^{-2\pi})^{\delta_{h_A 0} +\delta_{\bar{h}_A 0}} \equiv \Lambda_A.
$$ 

The derivation now proceeds as in \cite{hell} (with the appropriate definitions). We take the ratio of the $p=3$ and $p=1$ expressions to get
\begin{equation}
\frac{\sum_{A=1}^{\infty}f_3(\Delta_A+\hat{E}_0) \Lambda_A \, \text{exp}(-2\pi\Delta_A)}{\sum_{B=1}^{\infty}f_1(\Delta_B+\hat{E}_0) \Lambda_B \, \text{exp}(-2\pi\Delta_B)}     =    \frac{b_3(\hat{E}_0)}{b_1(\hat{E}_0)}\equiv F_0(\hat{E}_0).
\end{equation}
Rearranging this expression gives
\begin{equation}
\frac{\sum_{A=1}^{\infty}   \left[ f_3(\Delta_A+\hat{E}_0)  - F_0(\hat{E}_0) f_1(\Delta_A+\hat{E}_0)    \right]     \Lambda_A \,  \text{exp}(-2\pi\Delta_A)}{\sum_{B=1}^{\infty}f_1(\Delta_B+\hat{E}_0)  \Lambda_B \, \text{exp}(-2\pi\Delta_B)}     =   0. \label{eq:hellratio}
\end{equation}
To proceed, we make several convenient definitions. The bracketed expression in the numerator is our quantity of interest. It is a polynomial in $\Delta_A$ (again, actually $h_A$ and $\bar{h}_A$), so we define it as $P_1(\Delta_A)$. We define ${\Delta}^+_{1}$ as the largest root of $P_1$ when $\Lambda_A = 1$, ${\tilde{\Delta}}^+_{1}$ as the largest root when $\Lambda_A\neq 1$, and $f_1^+$ to be the largest root of $f_1$.

We perform a proof by contradiction and assume $\Delta_1 > \mbox{max}(\Delta_{1}^+,\tilde{\Delta}^+_{1}, f_1^+)$. Because $\Delta_A \geq \Delta_1$, our assumption implies that every term in both the numerator and denominator is strictly positive. Then equation (\ref{eq:hellratio})  says that a sum of positive numbers equals zero --- an impossibility. We conclude our assumption was incorrect so that
\begin{equation}
\Delta_1 \leq \mbox{max}(\Delta_1^+,\tilde{\Delta}^+_{1} ,f_1^+). \label{eq:genbound1}
\end{equation}
From the explicit form of $f_1(\Delta+\hat{E}_0$) in (2.12), we see that 
\begin{equation}
f_1^+ = \frac{c_{\rm tot}}{24}+\frac{(3-\pi)}{12\pi}. \label{eq:delfp}
\end{equation}
From \cite{hell}, we know that $\Delta_1^+$ is bounded above by
\begin{equation}
\Delta^+_1 \leq \frac{c_{\rm tot}}{12} + \delta_0 \approx \frac{c_{\rm tot}}{12} +.4736...
\end{equation}
We will now turn our attention to deriving a manageable expression for $\tilde{\Delta}^+_1$.
 
\subsection{Asymptotic and analytic bound on $\Delta_1$}

Following \cite{hell}, we begin by considering the limit of large positive total central charge $c_{\rm tot}$. In the limit $c_{\rm tot}\rightarrow\infty$, $\tilde{\Delta}^+_{1}$ is proportional to $c_{\rm tot}$, plus corrections of order $c^0_{\rm tot}$. We thus expand $\tilde{\Delta}^+_{1}$ as a series at large central charge:
\begin{equation}
\tilde{\Delta}^+_{1}\equiv\sum_{a=-1}^{\infty}d_{-a}\left( \frac{c_{\rm tot}}{24}\right)^{-a}, \,\, \mbox{such that  } P_1(\tilde{\Delta}^+_{1})=0. \label{eq:serdef}
\end{equation}
By definition, $\tilde{\Delta}^+_{1}$ is the largest real value with this property.
Substituting equation (\ref{eq:serdef}) into the explicit form of $P_1(\tilde{\Delta}^+_{1})=0$, the equation to leading order in $c_{\rm tot}$ is:
\begin{equation}
\frac{\pi^3}{1728} d_1(d_1-1)(d_1-2)=0.
\end{equation}
We choose $d_1=2$ so that $\tilde{\Delta}^+_{1}$ takes its largest value.

Solving to the next order in $c_{\rm tot}$, we find the expression
\begin{equation*}
 \frac{\pi^3}{36}  d_0  - \frac{\pi^3}{36}\frac{(\delta_{0h}+\delta_{0\bar{h}})}{e^{2\pi}-1} - \frac{\pi^2}{18} + \frac{\pi^3}{216} \frac{e^{2\pi}-13}{e^{2\pi}-1}=0.
\end{equation*}
\begin{equation*}
\Rightarrow d_0 = \frac{(12-\pi)e^{2\pi}-12+13\pi+6\pi}{6\pi(e^{2\pi}-1)} = \delta_0 + \frac{1}{e^{2\pi}-1} \approx 0.4755...
\end{equation*}
Thus we see that at this order, max$(\Delta_{1}^+,\tilde{\Delta}^+_{1},\Delta_{f_1}^+) =  \tilde{\Delta}^+_{1}$; for large enough central charge $c_{\rm tot}$, we can always bound the conformal dimension $\Delta_1$ using the expression
\begin{equation}
\Delta_1 \leq \frac{c_{\rm tot}}{12} + 0.4755... + O(c^{-1}_{\rm tot}). \label{eq:asymdel1}
\end{equation}

An absolute bound on $\Delta_1$ can be obtained with additonal work. Following steps similar to those in the appendices of \cite{hell}, we can show that a least upper linear bound on $\tilde{\Delta}^+_1$ is given by the first two terms of equation (\ref{eq:asymdel1}). From Appendix A.5 of \cite{hell}, we know that $\Delta^+_1$ is bounded above by $\frac{c_{\rm tot}}{12} + 0.4736...$ Thus the bound (\ref{eq:genbound1}) simplifies to
\begin{equation}
\Delta_1 \leq \frac{c_{\rm tot}}{12} + 0.4755...
\end{equation}
This is a universal bound, true for all 2d CFTs with modular invariance, discrete spectra, and $c, \bar{c} > 1$.



\section{Bounds on $\Delta_2, \Delta_n$}

In this section, we extend the methods of \cite{paper1} as above to derive bounds on primaries of second-lowest dimension. In order to bound the conformal dimension $\Delta_2$, we form the ratio of the $p=3$ and $p=1$ equations (\ref{eq:modeq}) beginning the sums now at $A,B=2$ to get
\begin{eqnarray}
\frac{\sum_{A=2}^{\infty}f_3(\Delta_A+\hat{E}_0) \Lambda_A e^{-2\pi\Delta_A}}{\sum_{B=2}^{\infty}f_1(\Delta_B+\hat{E}_0) \Lambda_B e^{-2\pi\Delta_B}} 	= \frac{f_3(\Delta_1+\hat{E}_0) \Lambda_1 e^{-2\pi\Delta_1}+b_3(\hat{E}_0)}{f_1(\Delta_1+\hat{E}_0) \Lambda_1 e^{-2\pi\Delta_1}+b_1(\hat{E}_0)} \equiv F_1(\Delta_1,c_{\rm tot}).  \label{eq:fdef}
\end{eqnarray}
Following work similar to Appendix $A$ of \cite{paper1}, we prove that $F_1$ is finite and nonzero for $0 < \Delta_1 \leq c_{\rm tot}/12 + .47...$ and $c,\bar{c}>1$. Subtracting $F_1$ over and simplifying gives the expression
\begin{equation}
\frac{\sum_{A=2}^{\infty} \left[f_3(\Delta_A+\hat{E}_0)-  f_1(\Delta_A+\hat{E}_0  ) F_1\right] \Lambda_A e^{-2\pi\Delta_A}}{\sum_{B=2}^{\infty}f_1(\Delta_B+\hat{E}_0) \Lambda_B e^{-2\pi\Delta_B}} =0  \label{eq:mine}
\end{equation}

In what follows, we define $f_p^+$ to be the largest root of $f_p(\Delta+\hat{E}_0)$ viewed as a polynomial in $\Delta$ (where once again the polynomial $f$ has additional dependence on $h$ and $\bar{h}$). The bracketed expression in the numerator is a polynomial cubic in $\Delta_A$; we denote this polynomial as $P_2(\Delta_A)$. We must be careful when defining the largest root of this polynomial. The largest root $\Delta^+_{2} (c_{\rm tot},\Delta_1)$ will change depending on the values of $h_1,\bar{h}_1,h_2,$ and $\bar{h}_2$. For now, we will keep things general and assume that the appropriate choices have been made in order to make this zero as large as possible.

With these definitions, we proceed via proof by contradiction. Following \cite{paper1}, we assume $\Delta_2>\text{max}(\Delta^+_{f_1},\Delta^+_{2} )$. As before, this assumption means every term in both the numerator and denominator of the left side of equation (\ref{eq:mine}) is positive. The left side thus can not be equal to zero, and we therefore have a contradiction.  Our assumption was incorrect, and we thus find
\begin{equation}
\Delta_2 \leq \text{ max}(\Delta^+_{f_1},\Delta^+_{2}). \label{eq:2bound}
\end{equation}

We can also obtain a numerical bound on $\Delta_2$ as in \cite{paper1}. We seek a linear bound of the form $\Delta_2 \leq \frac{c_{\rm tot}}{12}+D_1$, where $D_1$ is a numerical constant independent of $\Delta_1$. We want the smallest $D_1$ such that the inequality is valid for $c_{\rm tot} > 2$ and for all possible values of $\Delta_1$, $h_{1}$, $\bar{h}_{1}$, $h_{2}$, and $\bar{h}_{2}$. We can derive a universal bound by maximizing the function $\Delta^+_2-\frac{c_{\rm tot}}{12}$ as a function of all its dependences over their appropriate domains. This function attains a global maximum $D_1\approx 0.5531...$ (occurs when $c_{\rm tot} \approx 2$, $\Delta_1 \approx 0.2669..., \delta_{h_1 0} + \delta_{\bar{h}_{1} 0} = 1$, and  $\delta_{h_2 0} + \delta_{\bar{h}_2 0} = 1$). Therefore
\begin{equation}
\Delta_2 \leq \frac{c_{\rm tot}}{12} + 0.5531...
\end{equation}
This is a weaker bound on $\Delta_2$ than found in \cite{paper1}; this weaker result is expected given that we are now considering 2d CFTs with no restriction on the existence of primary operators.

Now that we have obtained a bound on $\Delta_2$, it is natural to extend our arguments to primary operators of higher dimension. A necessary condition that must hold for our method of proof to work for $\Delta_n$ is that $F_{n-1}$, defined as
\begin{equation}
F_{n-1}(\hat{E}_0,\Delta_1,\cdots,\Delta_{n-1}) \equiv \frac{\sum_{i=1}^{n-1} f_3(\Delta_i+\hat{E}_0)\Lambda_i\text{exp}(-2\pi\Delta_i)+b_3(\hat{E}_0)}{\sum_{i=1}^{n-1}f_1(\Delta_i+\hat{E}_0)\Lambda_i\text{exp}(-2\pi\Delta_i)+b_1(\hat{E}_0)},  \label{eq:fndef}
\end{equation}
be well-defined for all relevant values of its arguments. If the denominator of $F_{n-1}$ does not vanish, we can proceed as above to obtain a bound
\begin{equation}
\Delta_n \leq \text{max}(\Delta^+_{f_1},\Delta_n^+), 
\end{equation}  
where $\Delta^+_{f_1}$ is given by (\ref{eq:delfp}) and we define the largest real root $\Delta_{n}^+$ of the polynomial
\begin{equation}
P_n(\Delta) \equiv f_3(\Delta+\hat{E}_0)-f_1(\Delta+\hat{E}_0) F_{n-1}.
\end{equation}
Although $\Delta_n^+$ is a function of $c_{\rm tot}, \Delta_1, \cdots, \Delta_{n-1}$, we will typically suppress these dependences.

Following \cite{paper1}, we expect a bound of the same form as before: 
\begin{equation}
\Delta_n \leq \Delta_n^+ <  \frac{c_{\rm tot}}{12} + O(1).
\end{equation}
As in \cite{paper1}, however, the quantity $F_{n-1}$ can become undefined for $n\geq 4$. The largest value of the central charge for which denominator of $F_3$ vanishes, which we will call $c_{D3}^+$, lies within the range $c_{\rm tot} > 2$. Thus for $c_{\rm tot} < c_{D3}^+$,we cannot use this method to set a bound on $\Delta_4$. Similarly, the largest value of the central charge for which the numerator of $F_3$ vanishes, $c_{N3}$, also lies within our range of central charge.

The resolution to this issue mirrors the one given in \cite{paper1}: we further restrict the allowed values for the total central charge to $c_{\rm tot}\geq \mbox{max}(c^+_{Dn},c^+_{Nn})$. Because the polynomials $f_p$ and $b_p$ differ from their counterparts in \cite{paper1} by only some constant terms, it is unsurprising that we again find $c^+_{Dn} > c^+_{Nn}$. We must therefore solve for the value of the central charge $c_{Dn}$ which causes the denominator of (\ref{eq:fdef}) to vanish. In Appendix A, we explicitly show that the largest $c_{Dn}$, defined as $c^+_{Dn}$, occurs when its arguments $\Delta_i$ approach degeneracy. We use this result in Appendix B to show that
\begin{equation}
c^+_{Dn}\approx \frac{12}{\pi} W_0[A (n-1)] \sim \frac{12}{\pi}\ln(n),  \label{eq:dtor}
\end{equation} 
where $A\approx 0.3780...$ and $W_0$ is the primary branch of the Lambert-$W$ function. If we therefore require 
\begin{equation}
\log{(n)} \lesssim \frac{\pi c_{\rm tot}}{12} + O(1) \label{eq:nbound},
\end{equation}
then for asymptotically large central charge we obtain a bound on $\Delta_n$ going as
\begin{equation}
\Delta_n \leq \frac{c_{\rm tot}}{12} + O(1)\label{eq:deltanbound}.
\end{equation}

We can have similar difficulties with this ``$O(1)$'' term as in \cite{paper1}--it is only $O(1)$ with respect to $c_{\rm tot}$. If the $O(1)$ term goes as $ \log(n)$ or larger, for example, we could pick up contributions as large as $O(c_{\rm tot})$. The discussion and resolution of this potential issue is nearly identical to the case of non-chiral CFTs discussed in \cite{paper1}; we direct the reader to the analysis in Appendix D of this reference for complete details. In summary, we use an expression for the bound on the largest root of a cubic \cite{poly}. Using some assumptions, we can massage this bound into the form \ref{eq:deltanbound}.

Before continuing, we remark that our results have implications for gravity in $2+1$ dimensions via the AdS/CFT correspondence as in \cite{hell,paper1}. In the case of AdS$_3$/CFT$_2$, we follow \cite{witten} and make the identifications
\begin{equation}
c+\bar{c}=\frac{3L}{G_N}\;\;\; \mbox{and}\;\;\; E^{(rest)}=\frac{\Delta}{L},
\end{equation}
where $L$ is the AdS radius, $G_N$ is Newton's constant, $E^{(rest)}$ is the rest energy of an object in the bulk of AdS, and $\Delta$ is the conformal dimension of the corresponding boundary operator. 
Our bound then says that the dual gravitational theory must have massive states in the bulk (without boundary excitations) with rest energies $M_n=\Delta_n/L$ satisfying
\begin{equation}
M_n\leq M_n^+\equiv\frac{1}{L}\Delta^+_n |_{c_{\rm tot}=\frac{3L}{G_N}}.
\end{equation} 
That is, so long as $\log{(n)} \lesssim \frac{\pi c_{\rm tot}}{12} + O(1)$, we can derive a bound of the form
\begin{equation}
M \leq \frac{1}{4G_N} + O(L^{-1}).
\end{equation}

Another way of stating this result (first derived in \cite{paper1} for the case with no chiral Virasoro primaries and then later in \cite{hart} for the general case) is that there are $N>\sim \exp(\pi c_{\rm tot}/12)$ states satisfying the dimension bound
\begin{equation}
\Delta \leq \frac{c_{\rm tot}}{12} + O(1).
\end{equation}
Using our AdS/CFT dictionary, the logarithm of the number $N$ of states satisfying the mass bound
$$
M \leq \frac{1}{4G_N} + O(L^{-1})
$$
satisfies
\begin{equation}
\log N \geq \frac{\pi L}{4 G_N} + O(1)
\end{equation}
As in \cite{paper1}, this more general bound is consisten with the actual entropy of a spinless BTZ black hole \cite{bh1,bh2}.

\section{Testing the bounds}

Considering only 2d CFTs having $c,\bar{c}>1$ and no non-Virasoro chiral algebras is quite restrictive. As such, it can be difficult to find candidate theories for testing modular bootstrapping bounds. By extending previous results to theories with additional chiral primary operators, we can check bounds using conformal field theories with, for example, Ka\u{c}-Moody symmetry algebras. Here we will explicitly consider the $u(1)_k$ theories and $su(2)_k$ theories.

The $u(1)_k$ theories are readily found in any standard text on conformal field theory; we follow the notation and terminology of \cite{blumen}. The partition function of the $u(1)_k$ theory is given by
\begin{equation}
Z_{u(1)}^{(k)}(\tau,\bar{\tau}) =\sum_{m=-k+1}^k |\chi_m^{(k)}|^2,
\end{equation}
where
$$
\chi_m^{(k)} = \frac{\Theta_{m,k}(\tau)}{\eta(\tau)}\;\;\;\;\;\mbox{and}\;\;\;\;\;
\Theta_{m,k}(\tau) = \sum_{n\in\mathbb{Z} + \frac{m}{2k}} q^{kn^2},\;\;\; -k+1\leq m \leq k.
$$
This is the theory of one free boson compactified on a circle of radius $R=\sqrt{2k}$. This matches the fact that the character $\chi$ contributes 1 to the central charge. By expanding the character as a series in $q$, we also find that the highest weight state corresponding to the character $\chi_m^{(k)}$ has conformal dimension 
$$
h = \frac{m^2}{4k}.
$$

The explicit form of the partition function shows that there are no chiral primary operators in this theory. Of course, this theory should still satisfy our more general bound. To see this, we consider $|\chi_m^{(k)}|^2$ and calculate 
\begin{equation}
\Delta - \frac{c_{\rm tot}}{12} = \frac{3m^2-k}{6k}.
\end{equation}
The $m=0$ contribution corresponds to the vacuum; some thought convinces us that
\begin{equation}
\Delta_1 - \frac{c_{\rm tot}}{12}= \frac{3-k}{6k}.
\end{equation}
This expression is maximized when the level $k=1$. We have therefore found that 
\begin{equation}
\Delta_1 - \frac{c_{\rm tot}}{12} \leq \frac13 < 0.47...
\end{equation}
The details of the $k=1$ theory are well-understood; it is also straightforward to show that $\Delta_2$ for this theory satisfies our bound. And for $k>1$, the partition function contains $m=-1$ characters corresponding to a highest weight representation with 
$$
\Delta = \frac{(-1)^2}{4k} + \frac{(-1)^2}{4k} = \Delta_1.
$$
These CFTs \emph{must} therefore contain a primary state satisfying our bound on $\Delta_2$, and thus the $u(1)_k$ CFTs satisfy our bounds. 

A more interesting example is the $su(2)_k$ conformal field theories. For these theories, the characters are expressed in terms of the generalized $\Theta$-function
$$
\chi_\ell^{(k)} (\tau,z) =\frac{\Theta_{\ell+1,k+2}(\tau,z) - \Theta_{-\ell-1,k+2}(\tau,z) }{\Theta_{1,2}(\tau,z) - \Theta_{-1,2}(\tau,z)},
$$
where
$$
\Theta_{\ell,k}(\tau,z) \equiv \sum_{n \in \mathbb{Z}+\ell/2k } q^{kn^2} e^{-2\pi i n k z}
$$
and we must carefully take the limit $z\rightarrow 0$. Modular invariant partition functions are constructed from the characters
$$
Z_{su(2)}^{(k)}(\tau,\bar{\tau}) = \sum_{\ell,\ell'}\chi_\ell^{(k)}(\tau) M^{(k)}_{\ell\ell'}\bar{\chi}^{(k)}_{\ell'}(\bar{\tau}),
$$
where the matrix $M$ has non-negative integer entries. All matrices $M$ corresponding to consistent modular invariant theories have been found with their corresponding partition functions according to the A-D-E classification \cite{m2,m3}\footnote{This classification of $SU(2)_k$ invariants is related via string theory compactifications to the ADE classifications of singularities which are related in turn via Type IIA-heterotic string duality to the ADE classification of simple Lie algebras.}.
The central charge for CFTs with Ka\u{c}-Moody symmetry algebras is
\begin{equation}
c = \frac{g k}{k+C_g}
\end{equation}
where $g$ is the dimension of the corresponding Lie algebra and $C_g$ is the corresponding dual Coxeter number. For the algebra $su(N)$, the central charge is thus given by
$$
c = \frac{(N^2-1)k}{k+N}.
$$
For the current case, this becomes
\begin{equation}
c = \frac{3k}{k+2}.
\end{equation}

For the $su(2)_k$ symmetry algebra, the conformal weight of a highest weight state in the spin $\ell/2$ representation can be shown to equal
\begin{equation}
h_\ell = \frac{\ell(\ell+2)}{4(k+2)}.
\end{equation}
Depending on the specific A-D-E type of theory, the level $k$ and included characters will differ. We will only consider two cases here. First, the $A_{n+1}$ theories corresponding to level $k=n\geq 1$:
\begin{equation}
Z_{A_{n+1}} = \sum_{\ell=0}^n \bigg|\chi_\ell^{(k)} \bigg|^2.
\end{equation}
Looking at the characters that appear in this partition function, we see that
$$
\Delta_1 = \frac{1\cdot(1+2)}{4(k+2)} + \frac{1\cdot(1+2)}{4(k+2)} = \frac{3}{2(k+2)}.
$$
It follows that
$$
\Delta_1 - c_{\rm tot}/12 = \frac{5}{2(k+2)}-\frac12. 
$$
The RHS has its maximum value of 1/3 at level $k=1$. Thus the bound is satisfied
$$
\Delta_1 \leq \frac{c_{\rm tot}}{12}+\frac13 \leq \frac{c_{\rm tot}}{12} + .47...
$$

In a similar way, we can consider the $E_8$ theory at level $k=28$:
\begin{equation}
Z_{E_8} = \bigg|\chi_0^{(k)} +\chi_{10}^{(k)} + \chi_{18}^{(k)}+\chi_{28}^{(k)} \bigg|^2 + \bigg|\chi_{6}^{(k)}+\chi_{12}^{(k)}+\chi_{16}^{(k)}+\chi_{22}^{(k)} \bigg|^2
\end{equation}
We find that 
$$
\Delta_1 = \frac{6\cdot(6+2)}{4(k+2)} + \frac{6\cdot(6+2)}{4(k+2)} = \frac{24}{k+2}.
$$
Then
$$
\Delta_1 - \frac{c_{\rm tot}}{12} = \frac{48-k}{2(k+2)} = \frac13 \leq .47...
$$
Once again, the bound is satisfied. The partition function also contains characters for highest-weight states satisfying
$$
\Delta = \frac{10(12)}{4(28+2)} + \frac{0(2)}{4(28+2)} = 1,
$$
such that
$$
\Delta-c_{\rm tot}/12 = 8/15 = 0.533...
$$
We have thus also shown that this theory must contain a primary operator satisfying the $\Delta_2$ bound.

We also briefly discuss the $so(N)_1$ current algebra due to their ubiquity in superstring models. The algebra is realizable using $N$ real free fermions transforming in the vector representation of $SO(N)$. Of course, our formula for the central charge reproduces this fact to give $c=N/2$. We also know that the smallest nonvacuum conformal weight for the theory of a single free fermion is $h=1/16$. The theory of $N$ fermions will still have this smallest conformal weight. It trivially follows that
\begin{equation}
\Delta_1 = \frac18  \leq \frac{N}{12} + .47... = \frac{c_{\rm tot}}{12} + .47...
\end{equation}
Clearly such theories will satisfy our bounds.

There are many more candidate theories one can consider. An interesting new direction for investigating modular bootstrapping bounds comes from studying toroidal compactifications of free bosons on even, self-dual lattices. The ``even'' and ``self-dual'' properties are equivalent to modular invariance, and considering different numbers of bosons lets us consider different values of the central charge. By varying the lattice on which we compactify our theory, we can achieve different values for $\Delta_1$. This direction is interesting, but beyond the scope of this note; we study this topic in \cite{upcoming}.

\section*{Acknowledgments}

This work is partially supported by a University of Kentucky fellowship and by NSF $\#0855614$ and $\#1214341$, as well as by the National Science Council through the grant No.101-2112-M-002-027-MY3, Center for Theoretical Sciences at National Taiwan University, and Kenda Foundation. The author wishes to thank Alfred Shapere for his feedback and suggested revisions, as well as enlightening discussions in both the early and late stages of this manuscript.

\appendix
\section{Appendix A}

In this appendix, we consider the value of the central charge  $c_{\rm tot}$ causing the denominator of $F_{n-1}$ to vanish and prove that it is maximized when the conformal dimensions $\Delta_1,\cdots,\Delta_{n-1}$ approach degeneracy. The denominator of $F_{n-1}$ vanishes when
$$
\frac{\pi c_{Dn}}{12} -  \left(\frac{\pi}{6} - \frac12    \right) =  \frac{ \sum_{A=1}^{n-1}\left(2\pi \Delta_A -\frac{2\pi (\delta_{h_A0}+\delta_{\bar{h}_A 0})}{e^{2\pi}-1}  \right) \Lambda_A e^{-2\pi \Delta_A}    -2e^{-2\pi}(1-e^{-2\pi})  }{ \left( (1-e^{-2\pi})^2 + \sum_{A=1}^{n-1}\Lambda_A e^{-2\pi \Delta_A}  \right)  }
$$
With appropriate definitions, we can rewrite this expression as
$$
\hat{c} = \frac{\sum_{A=1}^{n-1}\left(2\pi \Delta_A -\frac{2\pi (\delta_{h_A0}+\delta_{\bar{h}_A 0})}{e^{2\pi}-1}  \right) \Lambda_A e^{-2\pi \Delta_A}+s_1}{\sum_{A=1}^{n-1} \Lambda_A e^{-2\pi \Delta_A} + s_2}\equiv \frac{N}{D}.
$$
In some of what follows, we will make use of the fact that $D>0$ for any values of its arguments. This is obvious from its can be seen from its explicit form.

In order for $\hat{c}$ to be a maximum when its arguments are identical, we need it to be a critical point and for the Hessian to be negative definite at this value (or equivalently, have all eigenvalues negative). We denote partial derivatives of $\hat{c}$ with respect to $\Delta_i$ as $\hat{c}_i$. We will need to calculate partial derivatives of $N$ or $D$ with respect to $\Delta_i$:
$$
N_i =2\pi \Lambda_i \exp(-2\pi\Delta_i) \left(1-\left(2\pi \Delta_i -\frac{2\pi (\delta_{h_i 0}+\delta_{\bar{h}_i 0})}{e^{2\pi}-1}  \right)  \right), \,\,\,\, N_{ij} = 0,
$$	
$$
N_{ii} = (2\pi)^2 \Lambda_i\exp(-2\pi\Delta_i) \left(-2 + \left(2\pi \Delta_i -\frac{2\pi (\delta_{h_i 0}+\delta_{\bar{h}_i 0})}{e^{2\pi}-1}  \right) \right)
$$
$$
D_i = -2\pi \Lambda_i \exp(-2\pi\Delta_i), \,\,\,\, D_{ij}= 0
$$
$$
D_{ii}= (2\pi)^2  \Lambda_i \exp (-2\pi\Delta_i).
$$
We then find
\begin{gather}
\hat{c}_i = \frac{N_i D - D_i N}{D^2} \nonumber \\
= \frac{2\pi \Lambda_i e^{-2\pi\Delta_1}}{D^2}
\bigg[     \left(1-\left(2\pi \Delta_i -\frac{2\pi (\delta_{h_i 0}+\delta_{\bar{h}_i 0})}{e^{2\pi}-1}  \right)   \right)\left(\sum_{A=1}^{n-1} \Lambda_A e^{-2\pi \Delta_A} + s_2 \right)   \\
  + \left(\sum_{A=1}^{n-1}2\pi \Delta_A \Lambda_A e^{-2\pi \Delta_A}+s_1\right) \bigg] . 
\end{gather}
The prefactor is nonvanishing. In order have a critical point, it is necessary and sufficient to have $\Delta$'s satisfying the condition
$$
2\pi\Delta_i^{crit.} + \delta_i = 1  +  \hat{c}( \Delta^{crit.}_1, \Delta^{crit.}_2, \cdots, \Delta^{crit.}_{n-1}),
$$
where to simplify our equations we have defined the value of $\Delta_j$ giving a critical point as $\Delta^{crit.}_j$ and  
$$
\delta_i \equiv - \left( \frac{2\pi (\delta_{h_i 0}+\delta_{\bar{h}_i 0})}{e^{2\pi}-1}  \right).
$$
The RHS of this equation will be the same for any value of $i$ on the LHS. This means that critical points will occur when $2\pi\Delta_1 + \delta_1 = 2\pi\Delta_2+\delta_2 = \cdots = 2\pi\Delta_{n-1} + \delta_{n-1}$. We will make use of this in detail in Appendix E.

To determine if this critical point is a maximum, we consider the Hessian. We calculate
$$
\hat{c}_{ii} = \frac{(N_i D - D_i N)_i D^2 - 2 D D_i  (N_i D - D_i N)}{D^4}.
$$
For critical points, the second term vanishes giving
$$
\hat{c}_{ii} \rightarrow \frac{(N_{ii} D - D_{ii} N) }{D^2}  = 
$$
$$\frac{(2\pi)^2 \Lambda_i e^{-2\pi\Delta_i}}{D^2}
\left[
(-2 + 2\pi\Delta_i + \delta_i) \left(\sum_{A=1}^{n-1} \Lambda_A e^{-2\pi \Delta_A} + s_2\right) - \left(\sum_{A=1}^{n-1} (2\pi \Delta_A + \delta_A) \Lambda_A e^{-2\pi \Delta_A}+s_1\right)  \right].
$$
Using our above condition for a critical point simplifies this expression to 
$$
\hat{c}_{ii} = - \frac{(2\pi)^2 \Lambda_i e^{-2\pi\Delta_i}}{D} < 0
$$
We will also need to calculate mixed partials:
$$
\hat{c}_{ij} = \frac{(N_i D - D_i N)_j D^2 - 2 D D_j (N_i D - D_i N)}{D^4},
$$
or in the case of a critical point
$$
\hat{c}_{ij} \rightarrow \frac{N_i D_j - D_i N_j }{D^2} = \frac{(2\pi)^2 \Lambda_i \Lambda_j e^{-2\pi\Delta_i} e^{-2\pi\Delta_j}}{D^2} (2\pi\Delta_i - 2\pi\Delta_j+\delta_i-\delta_j).
$$
Again using our condition for critical points, we see that all mixed partials will vanish. This means that the Hessian for the case where $2\pi\Delta_1 + \delta_1 = 2\pi\Delta_2+\delta_2 = \cdots = 2\pi\Delta_{n-1} + \delta_{n-1}$ is diagonal with purely negative entries; all eigenvalues are negative. Thus by our analysis we conclude that the function $\hat{c}$ (and thus $c_{Dn}$) will have a local maximum in the situation where all of its arguments are identical.

\section{Appendix B}

Here we will sketch the proof of the condition on $c_{\rm tot}$ given by equation (\ref{eq:dtor}). We begin with the condition that the denominator of $F_{n-1}$ vanishes and that this value is maximized when $2\pi\Delta_1 + \delta_1 = 2\pi\Delta_2+\delta_2 = \cdots = 2\pi\Delta_{n-1} + \delta_{n-1}$:
$$
\frac{\pi c_{Dn}}{12} -  \left(\frac{\pi}{6} - \frac12    \right) = \frac{\sum_{A=1}^{n-1}\left(2\pi \Delta_A + \delta_A  \right) \Lambda_A e^{-2\pi \Delta_A}+s_1}{\sum_{A=1}^{n-1} \Lambda_A e^{-2\pi \Delta_A} + s_2}
$$
$$
= \frac{ \left(2\pi \Delta_1 + \delta_1  \right)   \sum_{A=1}^{n-1}\Lambda_A e^{-2\pi \Delta_A}e^{-\delta_A}e^{\delta_A}  + s_1}{\sum_{A=1}^{n-1}\Lambda_A e^{-2\pi \Delta_A}e^{-\delta_A}e^{\delta_A} + s_2}
$$
$$
=\frac{ \left(2\pi \Delta_1 + \delta_1  \right) e^{-2\pi \Delta_1}e^{-\delta_1}  \sum_{A=1}^{n-1} \Lambda_A e^{\delta_A}  + s_1}{   e^{-2\pi \Delta_1}e^{-\delta_1}  \sum_{A=1}^{n-1} \Lambda_A e^{\delta_A} + s_2} = \frac{ \left(2\pi \Delta_1 + \delta_1  \right) e^{-2\pi \Delta_1}e^{-\delta_1}  m  + s_1}{   e^{-2\pi \Delta_1}e^{-\delta_1}  m + s_2} 
$$
We have defined
$$
\delta_A\equiv -\frac{2\pi (\delta_{h_A 0}+\delta_{\bar{h}_A 0})}{e^{2\pi}-1},    \,\,\,\,    s_1 \equiv -4\pi e^{-2\pi} (1-e^{-2\pi})
$$
$$
s_2 \equiv (1-e^{-2\pi})^2 ,    \,\,\,\,    m \equiv \sum_{A=1}^{n-1}\Lambda_A e^{\delta_A}
$$

The RHS will be maximized for
$$
\Delta_1 = \frac{1}{2\pi} W_0(m A) + \frac{B_1}{2\pi},
$$
with
$$
A \equiv \frac{e^{-\frac{s_1}{s_2} - 1}}{s_2}, \,\,\,\, B_1 \equiv \frac{s_1}{s_2} + 1 - \delta_1
$$
Substituting this back into our expression for the central charge, we find a complicated expression. We simplify it using the definition of the Lambert-$W$ function
$$
z = W_0(z) e^{W_0(z)} \rightarrow  e^{-W_0(z)} = \frac{W_0(z)}{z}.
$$
After some algebra, we find the largest value of the total central charge causing the denominator to vanish
\begin{equation}
\frac{\pi c^+_{Dn}}{12} = W_0(m A) + R_1 
\end{equation}
where
$$
R_1 \equiv \frac{-4\pi}{e^{2\pi}-1} + \left(\frac{\pi}{6} - \frac12    \right).
$$

Let us now turn our attention to the factor 
$$m\equiv \sum_{A=1}^{n-1} (1-e^{-2\pi})^{\delta_{h_A 0}+\delta_{\bar{h}_A 0}} \exp\left[-\frac{2\pi}{e^{2\pi}-1}     (\delta_{h_A 0}+\delta_{\bar{h}_A 0})  \right].$$
 How does a term in this sum contribute? If the Kronecker deltas vanish, then the argument of the sum is unity. On the other hand, if the sum of deltas is unity then the argument of the sum is approximately 0.9864. Since we have $(n-1)$ terms in the sum, this means that
$$
m = \alpha(n-1), \;\;\;\;\; 0.9864 \leq \alpha \leq 1.
$$

For large arguments of the Lambert-$W$ function, we can use the fact that $W_0(z) \approx \ln(z)$, plus $O(\ln(\ln(z)))$ corrections. For large enough $n$, the RHS will go as $\ln(n)$. We will restrict the total central charge so that $c_{\rm tot}>c_{Dn}^+$, meaning that to leading order we must require
\begin{equation}
c_{\rm tot} > \frac{12}{\pi} \ln(n).
\end{equation}
This is the result mentioned in the text. 

\section{Appendix C}
	
In this appendix, we provide an argument for the numerical observation that the bound on $\Delta_n$ for ranges we consider (for example, $W_0(n) < \left(c_{\rm tot} \right)^{1-\epsilon}$  and $\gamma \sim \left(c_{\rm tot}\right)^{1-\frac{\epsilon}{2}}$ with small $\epsilon$,) is maximized when the $\Delta_1, \Delta_2, \cdots, \Delta_{n-1}$ approach degeneracy. We will consider the case of theories like those found in \cite{hell}--- no chiral primary operators other than components of the stress tensor or their chiral descendants. The more general case follows in a nearly identical way, it is only more cumbersome.

To begin, we will show that nearly degenerate $\Delta$'s will maximize the function $F_{n-1}$. According to Appendix $D$ of \cite{paper1}, then, for the limits we consider here the function $\Delta_n^+-\frac{c_{\rm tot}}{12}$ will be take its maximum when $\sqrt{|F_{n-1}|}-\frac{c_{\rm tot}}{24}$ is maximized. Thus maximizing $F_{n-1}$ will maximize our bound. The quantity $F_2$ has degenerate $\Delta$'s trivially (as there is only $\Delta_1$). It can be shown analytically that for some value of $\Delta_1$, $F_2$ takes its maximum value. The conditions associated with this  are
$$
\frac{\partial}{\partial \Delta_1} F_2  \bigg|_{\Delta_1=\Delta_1^{max}} = 0
$$
$$
\Leftrightarrow \left( f_3'(\Delta_1^{max}+\hat{E}_0)-2\pi f_3(\Delta_1^{max}+\hat{E}_0)\right) \left( f_1(\Delta_1^{max}+\hat{E}_0)e^{-2\pi \Delta_1^{max}}+b_1(\hat{E}_0)    \right) 
$$
$$= \left( f_1'(\Delta_1^{max}+\hat{E}_0)-2\pi f_1(\Delta_1^{max}+\hat{E}_0)\right) \left( f_3(\Delta_1^{max}+\hat{E}_0)e^{-2\pi \Delta_1^{max}}+b_3(\hat{E}_0)    \right)
$$
and
$$
\frac{\partial^2}{\partial \Delta_1^2}  F_2  \bigg|_{\Delta_1=\Delta_1^{max}}  < 0
$$
$$
\Leftrightarrow  \left( f_3''(\Delta_1^{max}+\hat{E}_0)-4\pi f_3'(\Delta_1^{max}+\hat{E}_0)+4\pi^2 f_3(\Delta_1^{max}+\hat{E}_0)   \right) \left( f_1(\Delta_1^{max}+\hat{E}_0)e^{-2\pi \Delta_1^{max}}+b_1(\hat{E}_0)    \right)  
$$
$$
< \left(   f_1''(\Delta_1^{max}+\hat{E}_0)-4\pi f_1'(\Delta_1^{max}+\hat{E}_0)+4\pi^2 f_1(\Delta_1^{max}+\hat{E}_0  \right)   \left( f_3(\Delta_1^{max}+\hat{E}_0)e^{-2\pi \Delta_1^{max}}+b_3(\hat{E}_0)    \right).
$$

We will now assume that this fact is true for some finite number of $\Delta$'s and see the effect of adding of one more term:
\begin{equation}
F_{k+1} = \frac{f_3(\Delta_k+\hat{E}_0) e^{-2\pi \Delta_k}+N}{f_1(\Delta_k+\hat{E}_0)  e^{-2\pi \Delta_k}+D},
\end{equation}
where $N$ and $D$ are the numerator and denominator respectively of $F_k$.
To see that degenerate $\Delta$'s maximize this function, we must check several conditions. The condition that the first derivative with respect to $\Delta_k$ vanishes means
$$\left( f_3'(\Delta_k^{max}+\hat{E}_0)-2\pi f_3(\Delta_k^{max}+\hat{E}_0)\right) \left( f_1(\Delta_k^{max}+\hat{E}_0)e^{-2\pi \Delta_k^{max}}+ D^{max}    \right) 
$$
$$= \left( f_1'(\Delta_k^{max}+\hat{E}_0)-2\pi f_k(\Delta_1^{max}+\hat{E}_0)\right) \left( f_3(\Delta_k^{max}+\hat{E}_0)e^{-2\pi \Delta_k^{max}}+N^{max}    \right),
$$
where $N^{max}$ and $D^{max}$ are evaluated at the critical point values. Note that the condition for a vanishing first derivative with respect to any of the other $\Delta$'s looks the same except we substitute $\Delta_i^{max}$ in place of $\Delta_k^{max}$. This condition is of the same form as for $F_2$, where we know a solution exists. In the case where $\Delta$'s are degenerate, it reduces to the case of $F_2$ differing only by the presence of factors of $(n-1)$. A solution can be found to this equation. Thus the case of degenerate $\Delta$'s corresponds to a critical point.

To ensure this point is a maximum, we need to consider the second derivatives. We will consider first the case of mixed partials. Taking derivatives of $F_{k+1}$ with respect to $\Delta_i$ and $\Delta_j$ (including with respect to $\Delta_k$) gives (suppressing $\hat{E}_0$)
$$
\frac{\partial^2}{\partial \Delta_i \partial \Delta_j}F_{k+1}\bigg|_{\{\Delta\}=\{\Delta^{max} \}} = \frac{e^{-2\pi \Delta_i} e^{-2\pi \Delta_j}}{\left(  f_1(\Delta_k^{max})+D^{max} \right)^2  } \,\,\, \times
$$
$$
[\,\,\left(  \partial_i f_3(\Delta_i^{max}) -2 \pi f_3(\Delta_i^{max})   \right) \left(  \partial_j f_1(\Delta_j^{max}) -2 \pi f_1(\Delta_j^{max})   \right)
$$
$$
- \left(  \partial_j f_3(\Delta_j^{max}) -2 \pi f_3(\Delta_j^{max})   \right) \left(  \partial_i f_1(\Delta_i^{max}) -2 \pi f_1(\Delta_i^{max})   \right)  \,\,].
$$
Clearly for degenerate $\Delta$'s, all of the mixed partials will vanish. The expression for a second derivative with respect a particular $\Delta$ (again suppressing $\hat{E}_0$) is 
$$
\frac{\partial^2}{\partial \Delta_i^2}  F_{k+}  \bigg|_{\{\Delta\}=\{\Delta^{max} \}}   
= \frac{e^{-2\pi\Delta_i}}{\left( f_1(\Delta_k^{max})e^{-2\pi \Delta_k^{max}} + D^{max}   \right)^2} \, \times
$$
$$[\,\,  \left( f_3''(\Delta_i^{max})-4\pi f_3'(\Delta_i^{max})+4\pi^2 f_3(\Delta_i^{max})   \right) \left( f_1(\Delta_k^{max})e^{-2\pi \Delta_k^{max}} + D^{max}    \right)  
$$
$$
- \left( f_1''(\Delta_i^{max})-4\pi f_1'(\Delta_i^{max})+4\pi^2 f_1(\Delta_i^{max})   \right) \left( f_3(\Delta_k^{max})e^{-2\pi \Delta_k^{max}} + N^{max}    \right) \,\,].
$$
The bracketed expression is once more of the same form as the condition necessary for $F_2$. In the case of degenerate $\Delta$'s, the expressions become identical save for the presence of some $(n-1)$ factors. And it can be shown in a similar way that this expression is strictly negative.

Thus for the case of degenerate $\Delta$'s, the second derivative test shows that $F_{k+1}$ has a local maximum. By the discussion in Appendix $D$ of \cite{paper1}, this corresponds to when $\sqrt{|F_{n-1}|}-\frac{c_{\rm tot}}{24}$ is maximized and thus in the limits we consider when the least upper linear bound $\Delta_n^+-\frac{c_{\rm tot}}{12}$ is extremized.

\end{document}